\documentclass[12pt]{article}
\usepackage{graphicx,epsfig,psfig,color,latexsym}
\newcommand{\bc}{\begin{center}}
\newcommand{\ec}{\end{center}}
\newcommand{\be}{\begin{equation}}
\newcommand{\ee}{\end{equation}}
\newcommand{\bea}{\begin{eqnarray}}
\newcommand{\eea}{\end{eqnarray}}
\newcommand{\ba}{\begin{array}}
\newcommand{\ea}{\end{array}}
\newcommand{\lb}{\label}
\newcommand{\rf}{\ref}
\newcommand{\bfg}{\begin{figure}[htbp]}
\newcommand{\efg}{\end{figure}}

\newcommand{\pr}{Phys. Rev. }
\newcommand{\np}{Nucl. Phys. }

\newcommand{\prp}{Phys. Rep. }

\newcommand{\pl}{Phys. Lett. }

\newcommand{\nc}{Nuovo Cimento }

\newcommand{\zp}{Z. Phys. }

\begin{document}

\begin{flushright}
IPNO-DR-04-10, LPT-04-135
\end{flushright}
\vspace{0.5 cm}
\bc
{\large \textbf{Quarkonium bound state equation \protect \\
in the Wilson approach with minimal surfaces\footnote{Talk given 
by H.S. at the Workshop Hadron Structure and QCD, Repino,
St. Petersburg, Russia, 18-22 May 2004.}}}
\vspace{1. cm}

F. Jugeau$^a$ and  H. Sazdjian$^b$\\
\vspace{0.25 cm}
\textit{$^a$Laboratoire de Physique Th\'eorique\footnote{Unit\'e
Mixte de Recherche 8627.},\\
Universit\'e Paris XI, B\^at. 210, F-91405 Orsay Cedex, France\\
\footnotesize{E-mail: Frederic.Jugeau@th.u-psud.fr}\\
\vspace{0.25 cm}
\normalsize
$^b$Groupe de Physique Th\'eorique,
Institut de Physique Nucl\'eaire\footnote{Unit\'e Mixte de Recherche
8608.},\\
Universit\'e Paris XI, F-91406 Orsay Cedex, France\\
\footnotesize{E-mail: sazdjian@ipno.in2p3.fr}}
\ec
\par
\renewcommand{\thefootnote}{\fnsymbol{footnote}}
\vspace{0.75 cm}

\bc
{\large Abstract}
\ec
\par
Wilson loop averages are evaluated for large contours and
in the large-$N_c$ limit by means of minimal surfaces. 
This allows the study of the quark-antiquark gauge invariant
Green function through its dependence on Wilson loops.
A covariant bound state equation is derived which in the 
center-of-mass frame and at equal-times takes the form of
a Breit--Salpeter type equation. The interaction potentials
reduce in the static case to a confining linear vector
potential. For moving quarks, flux tube like contributions are
present. The nonrelativistic limit is considered. 
\par   
\vspace{0.5 cm}
\par
\newpage

The Wilson loop approach \cite{w} is based on the use of the 
gluon field path-ordered phase factor along a line $C$ joining
a point $x$ to a point $y$:
\be \lb{e1}
U(C_{yx},y,x)\equiv U(y,x)=Pe^{{\displaystyle -ig\int_x^y 
dz^{\mu}A_{\mu}(z)}}.
\ee
Equations satisfied by phase factors were obtained and analyzed by 
Mandelstam \cite{m1} and Nambu \cite{nm}.
\par
The Wilson loop is defined as the trace in color space of the 
phase factor on a closed contour $C$:
\be \lb{e2}
\Phi(C)=\frac{1}{N_c}\mathrm{tr}Pe^{{\displaystyle 
-ig\oint_Cdx^{\mu}A_{\mu}(x)}}.
\ee
Its vacuum expectation value is denoted $W(C)$:
\be \lb{e3}
W(C)=\langle\Phi(C)\rangle.
\ee
Loop equations were obtained by Polyakov \cite{p} and Makeenko and 
Migdal \cite{mm1,mm2,mgd,mk}. The Wilson loop essentially satisfies 
two types of equation, which are equivalent to the QCD
equations of motion:
\par
1) The Bianchi identity.
\par
2) The loop equations (or Makeenko--Migdal equations). Those actually
represent an infinite chain of coupled equations.
\par
A third property, factorization, is obtained in the large-$N_c$ limit
for two disjoint contours:
\be \lb{e4}
W(C_1,C_2)=W(C_1)W(C_2).
\ee
\par
Considerable simplification occurs in the large-$N_c$ limit of
the theory \cite{th}. In that limit, for large contours, i.e., at large
distances, nonperturbative asymptotic solutions to the Wilson
loops are represented by the minimal surfaces having as supports
the loop contours \cite{mm1,mm2}. Among various types of surfaces, 
minimal surfaces are the only ones that satisfy the Bianchi identity;
actually the latter becomes identical to the defining equation of
minimal surfaces \cite{js}. Furthermore, if at large distances
one ignores short-distance perturbative solutions, then the 
unrenormalized coupling constant which appears in the loop equations
can be adjusted in relation with the string tension \cite{js} and
therefore minimal surfaces define on their own independent solutions.
That property fails, however, when short-distance perturbative
solutions are taken into account, since in this case the unrenormalized
coupling constant is adjusted with respect to the perturbative regime.
Minimal surfaces then become only large-distance asymptotic solutions.
Nevertheless, if one is interested only in the large-distance behavior
of the theory, saturation of the Wilson loop averages by minimal
surfaces provides a correct description of the theory in this regime.
In that case, the Wilson loop average can be represented by the
following functional of the contour $C$:
\be \lb{e5}
W(C)=e^{{\displaystyle -i\sigma A(C)}},
\ee
where $\sigma$ is the string tension and $A(C)$ the minimal area
with contour $C$.
\par
\par 
Minimal surfaces also appear as natural solutions to the Wilson
loop averages in two-dimensional gauge theories \cite{kkk}. 
\par
To deal with the quarkonium bound state problem, we start 
with the two-particle gauge invariant Green function for
quarks $q_1$ and $q_2$ with different flavors and with masses 
$m_1$ and $m_2$:
\be \lb{e6}
G(x_1,x_2;x_1',x_2')\equiv \langle \overline \psi_2(x_2)U(x_2,x_1)
\psi_1(x_1)\overline \psi_1(x_1')U(x_1',x_2')\psi_2(x_2')
\rangle_{A,q_1,q_2}.
\ee
Here $U(x_2,x_1)$ is the phase factor (\rf{e1}) taken along the 
straight-line $x_1x_2$ (and similarly for $U(x_1',x_2')$).
Integrating in the large-$N_c$ limit with respect to the
quark fields, one obtains:
\be \lb{e7}
G(x_1,x_2;x_1',x_2')=-\langle \mathrm{tr}\,U(x_2,x_1)S_1(x_1,x_1')
U(x_1',x_2')S_2(x_2',x_2)\rangle_A,
\ee
where $S_1$ and $S_2$ are the quark and antiquark propagators in the 
presence of the external gluon field and $\mathrm{tr}$ designates the 
trace with respect to the color group.
\par
The Green function $G$ satisfies the following equation with respect to
the Dirac operator of particle 1 acting on $x_1$:
\bea \lb{e8}
& &(i\gamma.\partial_{(x_1)}-m_1)G(x_1,x_2;x_1',x_2')=
-i\langle\mathrm{tr}\,U(x_2,x_1)\delta^4(x_1-x_1')U(x_1',x_2')
S_2(x_2',x_2)\rangle_A\nonumber \\
& &\ \ \ \ \ \ -i\gamma^{\alpha}\langle\mathrm{tr}\int_0^1 
d\sigma(1-\sigma)\frac{\delta U(x_2,x_1)}{\delta x^{\alpha}(\sigma)}
S_1(x_1,x_1')U(x_1',x_2')S_2(x_2',x_2)\rangle_A,
\eea
where the segment $x_1x_2$ has been parametrized with the parameter
$\sigma$ as $x(\sigma)=(1-\sigma)x_1+\sigma x_2$; furthermore, the
operator $\delta/\delta x^{\alpha}$ does not act on the explicit
boundary point $x_1$  of the segment, this contribution having  
been cancelled by the contribution of the gluon field $A$ coming from 
the quark propagator $S_1$. A similar equation also holds with the Dirac 
operator of particle 2 acting on $x_2$. The operation
$\delta U(x_2,x_1)/\delta x(\sigma)$ introduces an insertion of
the field strength $F$ at the point $x(\sigma)$ of the 
straight-line $x_1x_2$ \cite{m1,p,mk}.
\par
In order to make apparent the Wilson loop structure of the two-particle
Green function, we adopt a representation for the quark propagator
in the external gluon field based on an explicit use of the phase
factor along straight lines \cite{js}. Introducing the gauge
covariant composite object $\widetilde S(x,x')$, made of a free fermion 
propagator $S_0(x-x')$ (without color group content) multiplied by the 
path-ordered phase factor $U(x,x')$ [Eq. (\rf{e1})] taken along the 
straight segment $x'x$,
\be \lb{e9}
\widetilde S(x,x')\equiv S_0(x-x')U(x,x'),
\ee
one shows that the quark propagator $S(x,x')$ in the external
gluon field satisfies the following functional integral equation in 
terms of $\widetilde S$:
\be \lb{e10}
S(x,x')=\widetilde S(x,x')-\int d^4x''S(x,x'')\gamma^{\alpha}
\int_0^1d\lambda\,\lambda\frac{\delta}{\delta x^{\alpha}(\lambda)}
\widetilde S(x'',x'),
\ee
where the operator $\delta/\delta x^{\alpha}(\lambda)$ acts on the
factor $U$ of $\widetilde S$, along the internal part of the segment 
$x'x''$, with $x'$ held fixed. A similar equation in which
the roles of $x$ and $x'$ are interchanged also holds. Those equations 
lead to iteration series for $S$ in which the gauge covariance property
is maintained at each order of the iteration. 
\par
Using the above representations for the quark propagators in Eq.
(\rf{e7}) one obtains for the two-particle Green function a series
expansion where each term contains a Wilson loop along a skew-polygon:
\be \lb{e11}
G=\sum_{i,j=1}^{\infty}G_{i,j},
\ee
where $G_{i,j}$ represents the contribution of the term of the series
having $(i-1)$ points of integration between $x_1$ and $x_1'$ ($i$
segments) and $(j-1)$ points of integration between $x_2$ and $x_2'$
($j$ segments). We designate by $C_{i,j}$ the contour associated
with the term $G_{i,j}$. A typical configuration for the contour of
$G_{4,3}$ is represented in Fig. \rf{f4}.
\par
\bfg
\vspace*{0.5 cm}
\bc
\begin{picture}(0,0)%
\epsfig{file=f4.pstex}%
\end{picture}%
\setlength{\unitlength}{2960sp}%
\begingroup\makeatletter\ifx\SetFigFont\undefined%
\gdef\SetFigFont#1#2#3#4#5{%
  \reset@font\fontsize{#1}{#2pt}%
  \fontfamily{#3}\fontseries{#4}\fontshape{#5}%
  \selectfont}%
\fi\endgroup%
\begin{picture}(6249,3405)(2464,-4474)
\put(5626,-1261){\makebox(0,0)[lb]{\smash{\SetFigFont{11}{13.2}{\familydefault}{\mddefault}{\updefault}{\color[rgb]{0,0,0}$y_2$}%
}}}
\put(6301,-2011){\makebox(0,0)[lb]{\smash{\SetFigFont{11}{13.2}{\familydefault}{\mddefault}{\updefault}{\color[rgb]{0,0,0}$y_3$}%
}}}
\put(4426,-1411){\makebox(0,0)[lb]{\smash{\SetFigFont{11}{13.2}{\familydefault}{\mddefault}{\updefault}{\color[rgb]{0,0,0}$y_1$}%
}}}
\put(3151,-2086){\makebox(0,0)[lb]{\smash{\SetFigFont{11}{13.2}{\familydefault}{\mddefault}{\updefault}{\color[rgb]{0,0,0}$x_1$}%
}}}
\put(6976,-2011){\makebox(0,0)[lb]{\smash{\SetFigFont{11}{13.2}{\familydefault}{\mddefault}{\updefault}{\color[rgb]{0,0,0}$x_1'$}%
}}}
\put(4351,-2761){\makebox(0,0)[lb]{\smash{\SetFigFont{11}{13.2}{\familydefault}{\mddefault}{\updefault}{\color[rgb]{0,0,0}$A_{4,3}$}%
}}}
\put(6001,-3586){\makebox(0,0)[lb]{\smash{\SetFigFont{11}{13.2}{\familydefault}{\mddefault}{\updefault}{\color[rgb]{0,0,0}$z_2$}%
}}}
\put(3076,-3661){\makebox(0,0)[lb]{\smash{\SetFigFont{11}{13.2}{\familydefault}{\mddefault}{\updefault}{\color[rgb]{0,0,0}$x_2$}%
}}}
\put(4126,-4411){\makebox(0,0)[lb]{\smash{\SetFigFont{11}{13.2}{\familydefault}{\mddefault}{\updefault}{\color[rgb]{0,0,0}$z_1$}%
}}}
\put(7276,-3436){\makebox(0,0)[lb]{\smash{\SetFigFont{11}{13.2}{\familydefault}{\mddefault}{\updefault}{\color[rgb]{0,0,0}$x_2'$}%
}}}
\end{picture}

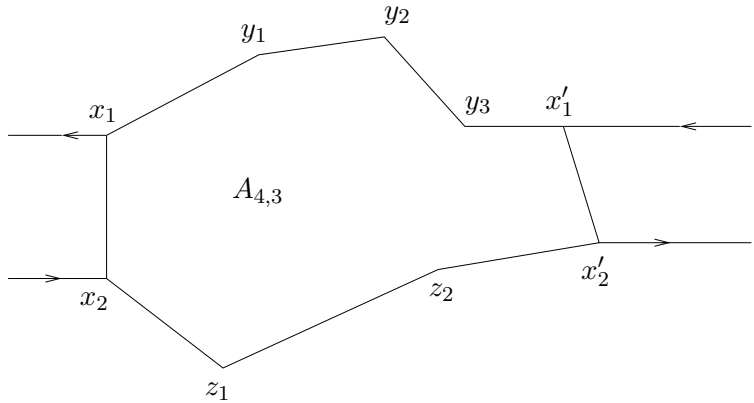
\caption{Contour $C_{4.3}$ associated with the term $G_{4,3}$.
$A_{4,3}$ is the minimal surface with contour $C_{4,3}$.}
\lb{f4}
\ec
\efg
Each segment of the quark lines supports a free quark propagator and 
except for the first segments (or the last ones, depending on the 
representation that is used) the Wilson loop is submitted to one 
functional derivative on each such segment. One then uses for the 
averages of the Wilson loops appearing in the above series the 
representation with minimal surfaces [Eq. (\rf{e5})].
\par
Representation (\rf{e10}) for the quark propagator is also used in 
Eq. (\rf{e8}) and its partner satisfied by the two-particle 
Green function $G$. One obtains two compatible equations for 
$G$ where the right-hand sides involve the series of the terms
$G_{i,j}$ of Eq. (\rf{e11}) and their functional derivative along
the segment $x_1x_2$. In order to obtain bound state equations,
it is necessary to reconstruct in the right-hand sides the bound
state poles contained in $G$ \cite{sb}. In $x$-space, bound states 
are reached by taking the large separation time limit between the pair 
of points ($x_1,x_2$) and ($x_1',x_2'$) \cite{gml}. 
\par
To produce a bound state pole, it is necessary that there be a coherent 
sum of contributions coming from each $G_{i,j}$, since the latter, 
taken individually, do not have poles. It is evident from
representation (\rf{e5}) that functional derivatives acting on
Wilson loop averages give rise to functional derivatives of the
corresponding minimal surfaces $A_{i,j}$. One then separates the various
contributions of the right-hand sides of Eq. (\rf{e8}) and of its
partner into different categories having the property of 
irreducibility. The first category corresponds
to terms in which the functional derivative along the segment 
$x_1x_2$ acts alone on a given $A$. In the second category, the
derivative along $x_1x_2$ is accompanied by another derivative 
along one of the segments of the quark lines, the two acting on
the same $A$. In the third category, the derivative along $x_1x_2$ 
and two other derivatives along segments of the quark lines act
on the same $A$ or on the product of two $A$s, respecting the
irreducibility property of kernels, in the sense that they are 
not parts of the series expansion of a factorized $G$, and so forth.
\par
Let us now consider the terms of the first category. One notices 
that the derivative along the segment $x_1x_2$ acts on areas 
$A_{i,j}$ with contour $C_{i,j}$ which are different from one term 
of the series to the other (the number of segments being different). 
To have a coherent sum of those contributions it is necessary to 
expand each such derivative around the derivative of the lowest-order 
contour $C_{1,1}$, represented in Fig. \rf{f1}.
\par
\bfg
\vspace*{0.5 cm}
\bc
\begin{picture}(0,0)%
\epsfig{file=f1.pstex}%
\end{picture}%
\setlength{\unitlength}{2960sp}%
\begingroup\makeatletter\ifx\SetFigFont\undefined%
\gdef\SetFigFont#1#2#3#4#5{%
  \reset@font\fontsize{#1}{#2pt}%
  \fontfamily{#3}\fontseries{#4}\fontshape{#5}%
  \selectfont}%
\fi\endgroup%
\begin{picture}(4662,2406)(2326,-3949)
\put(4201,-2761){\makebox(0,0)[lb]{\smash{\SetFigFont{11}{13.2}{\familydefault}{\mddefault}{\updefault}{\color[rgb]{0,0,0}$A_{1,1}$}%
}}}
\put(2326,-3661){\makebox(0,0)[lb]{\smash{\SetFigFont{11}{13.2}{\familydefault}{\mddefault}{\updefault}{\color[rgb]{0,0,0}$x_2$}%
}}}
\put(6901,-3886){\makebox(0,0)[lb]{\smash{\SetFigFont{11}{13.2}{\familydefault}{\mddefault}{\updefault}{\color[rgb]{0,0,0}$x_2'$}%
}}}
\put(6451,-1711){\makebox(0,0)[lb]{\smash{\SetFigFont{11}{13.2}{\familydefault}{\mddefault}{\updefault}{\color[rgb]{0,0,0}$x_1'$}%
}}}
\put(2476,-2011){\makebox(0,0)[lb]{\smash{\SetFigFont{11}{13.2}{\familydefault}{\mddefault}{\updefault}{\color[rgb]{0,0,0}$x_1$}%
}}}
\end{picture}

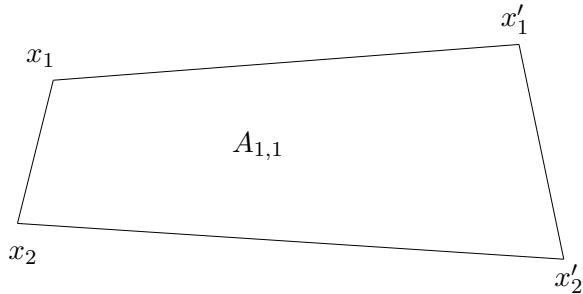
\caption{The lowest-order contour $C_{1,1}$ and its minimal surface
$A_{1,1}$.}
\lb{f1}
\ec
\efg
It is that term that can be factorized and can lead through the
summation of the factored series to the reappearance of the
Green function $G$ and to its poles. The remaining terms do not
lead to pole terms. Similarly, in the second category of terms,
the two derivative contributions containing the derivative along 
$x_1x_2$ should be expanded around the lowest-order contribution
coming from the contours $C_{2,1}$ or $C_{1,2}$, and so forth.
\par
In general, the derivative of the areas along $x_1x_2$ depends
among others on the slope of the areas in the orthogonal direction 
to $x_1x_2$, thus feeling the positions of the other points
$x_1'$ and $x_2'$. In order to obtain bound state equations
depending solely on $x_1$ and $x_2$ and not on the remote
points $x_1'$ and $x_2'$, it is necessary to associate the slopes
of the areas along the segment $x_1x_2$ with the quark momenta.
Taking then the large separation time limit, one ends up with two
covariant compatible bound state equations, in which the 
interaction kernels or potentials are given by various functional
derivatives involving at least one derivative along the segment 
$x_1x_2$. 
\par
The fact that one has two independent but compatible equations
means that the relative time variable does not play here a
dynamical role and could in principle be eliminated. That is done
by taking the difference and sum of the two equations. One finally
ends up with one single three-dimensional equation at equal-times 
in the center-of-mass frame, having the structure of a
Breit--Salpeter type equation \cite{b,s}. Keeping for the potentials 
the terms containing one functional derivative of the area $A_{1,1}$
[Fig. \rf{f1}], the equation takes the form \cite{js} 
\be \lb{e12}
\Big[P_0-(h_{10}+h_{20})-\gamma_{10}\gamma_1^{\mu}A_{1\mu}
-\gamma_{20}\gamma_2^{\mu}A_{2\mu}\Big]\psi(\mathbf{x})=0,
\ee
where $\psi$ is a $4\times 4$ matrix wave function of the relative 
coordinate $x=x_2-x_1$ considered at equal times, $P_0$ the 
center-of-mass total energy and $h_{10}$ and $h_{20}$ the quark and 
antiquark Dirac hamiltonians; the Dirac matrices of the quark (with 
index 1) act on $\psi$ from the left, while the Dirac matrices of the 
antiquark (with index 2) act on $\psi$ from the right. The potentials 
$A_1$ and $A_2$ are defined through the equations 
\be \lb{e13}
A_{1\mu}=\sigma\int_0^1d\sigma'(1-\sigma')\frac{\delta A_{1,1}}
{\delta x^{\mu}(\sigma')},\ \ \ \ \ 
A_{2\mu}=\sigma\int_0^1d\sigma'\,\sigma'\frac{\delta A_{1,1}}
{\delta x^{\mu}(\sigma')},
\ee
$x(\sigma')$ belonging to the segment $x_1x_2$.
\par
The time components of $A_1$ and $A_2$ add up in the wave
equation. For their sum, one has the expression (in the c.m. frame)
\bea \lb{e14}
& &A_{10}+A_{20}=\sigma r\frac{E_1E_2}{E_1+E_2}
\bigg\{\Big(\frac{E_1}{E_1+E_2}\epsilon(p_{10})+
\frac{E_2}{E_1+E_2}\epsilon(p_{20})\Big)\nonumber \\
& &\ \ \ \ \ \ \ \ \times\sqrt{\frac{r^2}{\mathbf{L}^2}}
\left(\,\arcsin\Big(\frac{1}{E_2}\sqrt{\frac{\mathbf{L}^2}{r^2}}\Big)+
\arcsin\Big(\frac{1}{E_1}\sqrt{\frac{\mathbf{L}^2}{r^2}}\Big)\,\right)
\nonumber \\
& &+(\epsilon(p_{10})-\epsilon(p_{20}))
\Big(\frac{E_1E_2}{E_1+E_2}\Big)\Big(\frac{r^2}{\mathbf{L}^2}\Big)
\left(\,\sqrt{1-\frac{\mathbf{L}^2}{r^2E_2^2}}-
\sqrt{1-\frac{\mathbf{L}^2}{r^2E_1^2}}\,\right)\bigg\}.\nonumber \\
& &
\eea
Here, $r=\sqrt{\mathbf{x}^2}$,
$E_a=\sqrt{m_a^2+\mathbf{p}^2}$, $a=1,2$, with $m_a$ the quark
masses, $\mathbf{p}$ the c.m. momentum, 
$\mathbf{p}=(\mathbf{p}_2-\mathbf{p}_1)/2$, $\mathbf{L}$ the c.m.
orbital angular momentum, and $\epsilon(p_{10})$ and $\epsilon(p_{20})$
the energy sign operators of the free quark and the antiquark, 
respectively:
\be \lb{e15}
\epsilon(p_{a0})=\frac{h_{a0}}{E_a},\ \ \ \ \ a=1,2.
\ee
\par 
The space components of $A_1$ and $A_2$ are orthogonal to 
$\mathbf{x}$. The expression of $\mathbf{A}_1$ is (in the c.m. frame):
\bea \lb{e16}
\mathbf{A}_1&=&-\sigma r\frac{E_1E_2}{E_1+E_2}
\bigg\{\,\frac{r^2}{2\mathbf{L}^2}\frac{E_1E_2}{E_1+E_2}\mathbf{p}^t
\nonumber \\
& &\ \ \ \ \ \times\sqrt{\frac{r^2}{\mathbf{L}^2}}
\left(\,\arcsin\Big(\frac{1}{E_2}\sqrt{\frac{\mathbf{L}^2}{r^2}}\Big)+
\arcsin\Big(\frac{1}{E_1}\sqrt{\frac{\mathbf{L}^2}{r^2}}\Big)\,\right)
\nonumber \\
& &\ \ \ +\frac{1}{E_2}\mathbf{p}^t\Big(\frac{E_1E_2}{E_1+E_2}\Big)
\Big(\frac{r^2}{\mathbf{L}^2}\Big)
\left(\,\sqrt{1-\frac{\mathbf{L}^2}{r^2E_2^2}}-
\sqrt{1-\frac{\mathbf{L}^2}{r^2E_1^2}}\,\right)\nonumber \\
& &\ \ \ -\frac{1}{2}\mathbf{p}^t\Big(\frac{r^2}{\mathbf{L}^2}\Big)
\left(\,\frac{E_1}{E_1+E_2}\sqrt{1-
\frac{\mathbf{L}^2}{r^2E_2^2}}+\frac{E_2}{E_1+E_2}
\sqrt{1-\frac{\mathbf{L}^2}{r^2E_1^2}}\,\right)\,\bigg\}.
\eea
Here, $\mathbf{p}^t$ is the transverse part of $\mathbf{p}$ with
respect to $\mathbf{x}$:
\be \lb{e17}
\mathbf{p}^t=\mathbf{p}-\mathbf{x}\frac{1}{\mathbf{x}^2}
\mathbf{x}.\mathbf{p}.
\ee
The expression of $\mathbf{A}_2$ is obtained from that of 
$\mathbf{A}_1$ by an interchange in the latter of the indices 1 and 2 
and a change of sign of $\mathbf{p}^t$.  
\par
For sectors of quantum numbers where $\mathbf{L}^2=0$, the expressions 
of the potentials become:
\bea 
\lb{e18}
& &A_{10}+A_{20}=\frac{1}{2}(\epsilon(p_{10})+\epsilon(p_{20}))
\sigma r,\\
\lb{e19}
& &\mathbf{A}_1=-\frac{1}{E_1E_2}\Big(\frac{1}{3}(E_1+E_2)-
\frac{1}{2}E_1\Big)\mathbf{p}^t\sigma r,\nonumber \\
& &\mathbf{A}_2=+\frac{1}{E_1E_2}\Big(\frac{1}{3}(E_1+E_2)-
\frac{1}{2}E_2\Big)\mathbf{p}^t\sigma r.
\eea
\par
The potentials are generally momentum dependent operators and 
necessitate an appropriate ordering of terms.
\par
From the structure of the wave equation (\rf{e12}) and the expressions 
of the potentials, one deduces that the interaction is confining and 
of the vector type. However, compared to the conventional timelike
vector potential, it has additional pieces of terms contributing
to the orbital angular momentum dependent parts. A closer analysis
of those terms shows that they can be interpreted as being originated
from the moments of inertia of the segment $x_1x_2$ carrying a constant
linear energy density equal to the string tension. The interaction
potentials are therefore provided by the energy-momentum vector of the
segment joining the quark to the antiquark, in similarity with the
color flux tube picture of confinement. An analogous equation had
also been proposed by Olsson \textit{et al.} on the basis of a model
where the quarks are attached at the ends of a straight string or a
color flux tube \cite{ow,lcooowoo}. A similar conclusion had also been
reached by Brambilla, Prosperi \textit{et al.} on the basis of the
analysis of the relativistic corrections to the nonrelativistic limit
of the Wilson loop \cite{bmpbbp,bcpbmp,bp}.
\par
For heavy quarks, one can expand equation (\rf{e12}) around the
nonrelativistic limit. To order $1/c^2$, the hamiltonian becomes
(in the c.m. frame):
\bea \lb{e20}
H&=&\frac{\mathbf{p}^2}{2\mu}+\sigma r-\frac{2\hbar\sigma}{\pi}
\Big(\frac{1}{m_1}+\frac{1}{m_2}\Big)
-\frac{1}{8}\Big(\frac{1}{m_1^3}+\frac{1}{m_2^3}\Big)(\mathbf{p}^2)^2
+\frac{\hbar^2}{4}\Big(\frac{1}{m_1^2}+\frac{1}{m_2^2}\Big)
\frac{\sigma}{r}\nonumber \\
& &-\frac{\sigma}{6r}\Big(\frac{1}{m_1^2}+\frac{1}{m_2^2}-\frac{1}{m_1m_2}
\Big)(\mathbf{L}^2+2\hbar^2)
+\frac{\sigma}{2r}\Big(\frac{\mathbf{L}.\mathbf{s}_1}
{m_1^2}+\frac{\mathbf{L}.\mathbf{s}_2}{m_2^2}\Big)\nonumber \\
& &-\frac{2\sigma}{3r}\Big(\frac{1}{m_1^2}-\frac{1}{2m_1m_2}\Big)
\mathbf{L}.\mathbf{s}_1-\frac{2\sigma}{3r}\Big(\frac{1}{m_2^2}-
\frac{1}{2m_1m_2}\Big)\mathbf{L}.\mathbf{s}_2.
\eea
[$\mu=m_1m_2/(m_1+m_2)$, $\mathbf{s}_1$ and $\mathbf{s}_2$ are the spin
operators of the quark and of the antiquark.]
Several remarks can be made at this stage. First, the hamiltonian is
independent of spin-spin interactions. Second, purely orbital angular 
momentum dependent pieces (proportional to $\mathbf{L}^2$) are present, 
the origin of which is related to
the contribution to the rotational motion of the system of the moments of
inertia of the flux tube, represented by the straight segment joining the
quark to the antiquark. Those terms were also
obtained in Refs. \cite{bmpbbp} and \cite{ow,lcooowoo}. Third, 
two kinds of spin-orbit term are present. The first  
comes from the contribution of a conventional timelike
vector interaction represented by the potential $\sigma r$, which is the
dominant part of the combination $A_{10}+A_{20}$ [Eq. (\rf{e14})].
The second type comes from the contributions of the direct
interactions of the momentum of the flux tube with the quarks, represented
by the spacelike potentials $\mathbf{A}_1$ and $\mathbf{A}_2$ 
[Eqs. (\rf{e16})]. The latter terms induce negative signs to the 
spin-orbit couplings, in opposite direction to the former one, a feature 
which is also observed on phenomenological grounds for the large-distance
effects in fine splitting.
\par
The relativistic corrections to the interquark potential arising
from the Wilson loop were analyzed and evaluated in the literature by 
Eichten and Feinberg \cite{ef}, Gromes \cite{gr}, Brambilla, Prosperi
\textit{et al.} \cite{bmpbbp,bcpbmp,bp}, Brambilla, Pineda, Soto and 
Vairo \cite{pvbpsv}.
\par
The Wilson loop approach was also used for the study of quarkonium 
systems by Dosch, Simonov \textit{et al.} with the use of the
stochastic vacuum model \cite{dssdg}.
\par
In conclusion, the saturation of the Wilson loop averages in the 
large-$N_c$ limit by minimal surfaces provides a systematic tool for 
investigating the large-distance dynamics of quarkonium systems. 
A complete study of those systems necessitates the incorporation of 
the short-distance contributions, taken into account by perturbation 
theory.
\par

\vspace{0.25 cm}
\noindent
\textbf{Acknowledgements}: This work was supported in part for H.S. by 
the European Community network EURIDICE under contract No. 
HPRN-CT-2002-00311.
\par

\end{document}